# Surface Motion Relative to the Irregular Celestial Bodies

Yu Jiang[1, 2], Yun Zhang[2], Hexi Baoyin[2]

1. State Key Laboratory of Astronautic Dynamics, Xi'an Satellite Control Center, Xi'an 710043, China
2. School of Aerospace Engineering, Tsinghua University, Beijing 100084, China

Y. Jiang (✉) e-mail: jiangyu_xian_china@163.com (corresponding author)

**Abstract**. We study the motion and equilibria of the grains on the surface of the irregular celestial body (hereafter called irregular bodies). Motions for the grains on the smooth and unsmooth surfaces are discussed, respectively. The linearized equations of motion relative to a surface equilibrium point and its characteristic equations are presented. Considering the stick-slip effect, the damping forces and the spring forces for the grain are calculated, then the linearized equations of motion and the characteristic equations relative to the surface equilibrium points are derived. The number of non-degenerate surface equilibria is an even number. We compute the motion of a grain released above three different regions relative to the irregular asteroid 6489 Golevka, including the flat surface, the concave region, and the convex region. Following the grain release and initial bounce, three kinds of motions exist: the orbital motion, the impact motion and the surface motion. We find that the maximum height of the next hop may be bigger than the maximum height of the current hopping. We also used Monte Carlo simulations to calculate 100 grains' hopping motions, the results shows that the stable surface equilibria are on the concave region and flat surface of the asteroid.

## 1. Introduction

Previous work have studied the physics and chemistry character of the surface grain and dust, such as the dust grain's electrostatic and rotational ejection from the cometary nucleus (Oberc 1997) and the mineralogy and mineral chemistry of the asteroid's dust particles (Nakamura et al. 2011). The motion of grain and dust can be caused by the YORP effect (Fahnestock and Scheeres, 2009), the windmill effect acting on the surface of small bodies (Oberc 1997), the collision and gravitational reaccumulation of asteroid families and moonlets (Michel et al. 2001), and the disruption of rubble-pile asteroids (Asphaug et al. 1998; Jewitt et al. 2014). The disintegration of the asteroid P/2013 R3 generated more than 10 distinct components, a great number of small grains, and a comet-like dust tail (Jewitt et al. 2014). Since most asteroids and comets have irregular shapes (Lagerros 1997; Gutiérrez et al. 2001), these grains may move on the surface of the irregular celestial bodies. To understand the surface motion of grains on the irregular minor celestial bodies, the surface mechanical environment of these bodies needs to be investigated.

Some simple-shaped bodies are considered to help understand the surface motion on an irregular body, including an ellipsoid (Guibout and Scheeres 2003; Bellerose and Scheeres 2008; Bellerose et al. 2009) and a massive cube (Liu et al. 2013). Guibout and Scheeres (2003) discussed the existence and stability of equilibria on the surface of a rotating ellipsoid. Bellerose and Scheeres (2008) modeled the asteroid as an ellipsoid and analyzed the hopping on a flat surface. Bellerose et al. (2009) also supposed the asteroid as an ellipsoid and discussed the motion and control of surface exploration robots on asteroids. The surface motion of a vehicle travelling on the surface of irregular minor celestial bodies is linked to the study of surface grain motion. If a vehicle lands on the surface of a minor celestial body, its impact and hop will be affected by the irregular surface (Bellerose et al. 2009; Tardivel et al. 2014). Liu et al. (2013) investigated the surface motion on a rotating massive cube and computed the locations and eigenvalues of the surface equilibria.

The dynamic equation of motion as well as the existence and stability of equilibria on the smooth surface of the ellipsoid is studied by Guibout and Scheeres (2003). The Lagrangian (Guibout and Scheeres 2003) and Jacobian integral (Bellerose and Scheeres 2008; Liu et al. 2013) can be defined using the gravitational potential and the motion parameters. The linear stability of equilibria on the smooth surface of a cube can be computed by the eigenvalues of the equilibria (Liu et al. 2013). The non-degenerate equilibria around the asteroid can be classified into eight different cases (Jiang et al. 2014, 2016; Jiang 2015), while the parameter varies, the position and topological cases of the equilibria will change (Jiang et al. 2015a, b). The motion on the surface of the asteroid is different from the motion around the asteroid (Jiang and Baoyin 2014; Jiang et al. 2015a; Chanut et al. 2015); the motion on the surface of the asteroid need to include the gravitational force of the asteroid and the contact force between the asteroid and the grain.

Considering the surface as unsmooth and having an invariable friction coefficient (Liu et al. 2013), the equilibria also exist and have different stability from the equilibria of smooth surface. However, the friction phenomena on the minor celestial bodies' irregular surface has the stick-slip effect (Fahnestock and Scheeres, 2009; Das et al. 2015), and the grains may be charged (Oberc 1997). Considering the detailed gravity and geomorphology, Yu and Baoyin (2014) analyzed the motion of the asteroid surface vehicle and found that the most stable direction is the rotational pole direction, which can limit the natural trajectory of the vehicle after it lands on the asteroid. Yu and Baoyin (2015) presented a numerical method to simulate the motion of migrating grains on the surface of the asteroids.

This study seeks to consider the small celestial body's irregular shapes and irregularly gravitational field of the small celestial body to analyze the motion of grains on the surface of the irregular body. An identity about the eigenvalues of equilibria is derived, which implies that the number of non-degenerate surface equilibria of the grain is even and can only change in pairs. For the cases of the grain on the smooth and unsmooth surfaces, the linearized motion equations relative to the surface equilibria and the characteristic equations of the surface equilibria are derived, and the sufficient conditions for the linear stability of the surface equilibria are presented. The degenerate surface equilibria can disappear, or change to even number of non-degenerate surface equilibria, or change to an arbitrary number of degenerate surface equilibria, or change to even number of non-degenerate surface equilibria and arbitrary number of degenerate surface equilibria. However, the degenerate surface equilibria cannot change to odd number of non-degenerate surface equilibria.

The irregular asteroid 6489 Golevka is chosen as example in the calculation of surface equilibria and hopping motions of 100 grains released above three different regions (the flat surface, the concave region, and the convex region) using Monte Carlo simulations. The results show that grains released above the convex region need significant longer time before settling on the surface equilibrium location. On the opposite, grains released above the flat surface and the concave region have significantly shorter trajectories. Considering the friction force, the Monte Carlo simulations of hopping motions show that the stable surface equilibria are on the concave region and flat surface; there are no stable surface equilibria on the convex region of the surface.

## 2. Surface Motion on the Irregular Body

In this section, we investigate the motion of the grain which is always on the surface of the irregular body. While the grain is moving on the surface of the irregular body with the friction force, the grain will be in the state of equilibrium on the body's surface in the end.

### 2.1 Surface Motion on the Smooth Surface

Let us consider the gravitational fields of the irregular, and the surface of the irregular body to be smooth. The irregular gravitational field and the shape model of the irregular body can be modelled by the the polyhedral model (Werner 1994; Werner and Scheeres 1997) using data from radar observations (Stooke 2002). The body's gravitational potential (Werner and Scheeres 1997) can be calculated by

$$U(\mathbf{r}) = \frac{1}{2}G\sigma \sum_{e \in edges} \mathbf{r}_e \cdot \mathbf{E}_e \cdot \mathbf{r}_e \cdot L_e - \frac{1}{2}G\sigma \sum_{f \in faces} \mathbf{r}_f \cdot \mathbf{F}_f \cdot \mathbf{r}_f \cdot \omega_f, \qquad (1)$$

where $\mathbf{r}$ is the position vector of the grain in the body-fixed frame of the irregular body, $U(\mathbf{r})$ is the gravitational potential of the irregular body, G=$6.67 \times 10^{-11}$ m$^3$kg$^{-1}$s$^{-2}$ is the gravitational constant, $\sigma$ is the density of the irregular body; $\mathbf{r}_e$ and $\mathbf{r}_f$ are vectors from the field points to fixed points on the edge of face and the face, respectively; $\mathbf{E}_e$ and $\mathbf{F}_f$ are dyads that defined in terms of face- and edge-normal vectors to represent geometric parameters of edges and faces, respectively; $L_e$ is the integrate factor, $\omega_f$ is the signed solid angle.

The surface motion of the grain on the irregular body

$$\begin{cases} \ddot{\mathbf{r}} = -2\boldsymbol{\omega}\times\dot{\mathbf{r}} - \dot{\boldsymbol{\omega}}\times\mathbf{r} - \dfrac{\partial V(\mathbf{r})}{\partial \mathbf{r}} + \mu_L \nabla W \\ W(\mathbf{r}) = 0 \end{cases}, \quad (2)$$

where $\boldsymbol{\omega}$ is the rotational angular velocity vector of the irregular body, $\mu_L$ is the Lagrange multiplier, $V(\mathbf{r}) = -\dfrac{1}{2}(\boldsymbol{\omega}\times\mathbf{r})\cdot(\boldsymbol{\omega}\times\mathbf{r}) + U(\mathbf{r})$, $W(\mathbf{r})$ is the surface function of the irregular body. Using $W(\mathbf{r})$, we can calculate the normal direction of the surface, and thus the friction force. In general, the surface is irregular and can only be computed by the numerical method, such as the polyhedral model and the N-body model.

If the surface is smooth, the linearized equations of motion relative to the surface equilibrium point can be given by

$$\begin{aligned} \ddot{\xi} &= -2\omega_y \dot{\zeta} + 2\omega_z \dot{\eta} - V_{xx}\xi - V_{xy}\eta - V_{xz}\zeta \\ \ddot{\eta} &= -2\omega_z \dot{\xi} + 2\omega_x \dot{\zeta} - V_{xy}\xi - V_{yy}\eta - V_{yz}\zeta \\ \ddot{\zeta} &= -2\omega_x \dot{\eta} + 2\omega_y \dot{\xi} - V_{xz}\xi - V_{yz}\eta - V_{zz}\zeta \end{aligned}, \quad (3)$$

where $\mathbf{K}_V = \begin{pmatrix} V_{xx} & V_{xy} & V_{xz} \\ V_{xy} & V_{yy} & V_{yz} \\ V_{xz} & V_{yz} & V_{zz} \end{pmatrix}_L$ is the Hessian matrix of $V(\mathbf{r})$, $\xi = x - x_L$, $\eta = y - y_L$, $\zeta = z - z_L$. The characteristic equation of the surface equilibrium is

$$\begin{vmatrix} \lambda^2 + V_{xx} & -2\omega_z \lambda + V_{xy} & 2\omega_y \lambda + V_{xz} \\ 2\omega_z \lambda + V_{xy} & \lambda^2 + V_{yy} & -2\omega_x \lambda + V_{yz} \\ -2\omega_y \lambda + V_{xz} & 2\omega_x \lambda + V_{yz} & \lambda^2 + V_{zz} \end{vmatrix} = 0. \quad (4)$$

Eq .(4) is a sextic equation. Using the local coordinates on the surface of the irregular body, the characteristic equation of the surface equilibrium can be simplified into a quartic equation. The original point of the local coordinates is defined as the surface equilibrium; the X axis and the Y axis are in the tangential plane of the surface equilibrium, the Z axis is perpendicular to the tangential plane and points to the outside of the body. Denote the Hessian matrix of $V(\mathbf{r})$ in the tangential plane as $\begin{pmatrix} V_{XX} & V_{XY} \\ V_{XY} & V_{YY} \end{pmatrix}_L$, and denote the component of the irregular body's rotation velocity about the Z axis as $\omega_r$. Then the characteristic equation of the surface equilibrium is

$$\lambda^4 + \left(4\omega_r^2 + V_{XX} + V_{YY}\right)\lambda^2 + \left(V_{XX}V_{YY} - V_{XY}^2\right) = 0 \tag{5}$$

**2.2 Surface Motion on the Unsmooth Surface**

If the surface is unsmooth, let the unit vector of the normal direction be $\mathbf{N}_\perp$ and the unit vector of the friction's direction be $\mathbf{N}_{//} = \left[N_x, N_y, N_z\right]$, as shown in Figure 1. Then

$$\mathbf{N}_\perp = \frac{\left(\frac{\partial W}{\partial x}, \frac{\partial W}{\partial y}, \frac{\partial W}{\partial z}\right)}{\sqrt{\left(\frac{\partial W}{\partial x}\right)^2 + \left(\frac{\partial W}{\partial x}\right)^2 + \left(\frac{\partial W}{\partial z}\right)^2}}, \quad \mathbf{N}_{//} = \frac{\frac{\partial W}{\partial \mathbf{r}}}{\left|\frac{\partial W}{\partial \mathbf{r}}\right|}. \tag{6}$$

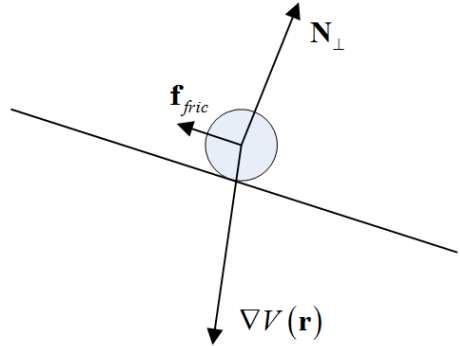

Figure 1. The normal direction and the friction force.

The pressure force is

$$\mathbf{N}_f = \left(\nabla U^* \cdot \mathbf{N}_\perp\right)\mathbf{N}_\perp, \tag{7}$$

where $\nabla U^* = \nabla V + 2m\boldsymbol{\omega} \times \dot{\mathbf{r}}$.

Denote $\Theta = \nabla U^* \cdot \mathbf{N}_\perp$, then the friction force is

$$\mathbf{f}_{fric} = \mu |\mathbf{N}_f| \mathbf{N}_{//} = \mu \Theta \mathbf{N}_{//}, \tag{8}$$

where $\mu$ is the coefficient of friction, which is invariant.

Eq. (2) can be rewritten as Eq. (9)

$$\begin{cases} \ddot{\mathbf{r}} = -2\boldsymbol{\omega} \times \dot{\mathbf{r}} - \dfrac{\partial V(\mathbf{r})}{\partial \mathbf{r}} - \mu \Theta \mathbf{N}_{//} + \mu_L \nabla W \\ W(\mathbf{r}) = 0 \end{cases}. \tag{9}$$

Let $\nabla V(\mathbf{r}) = -\omega^2 \mathbf{r} + \nabla U(\mathbf{r}) - \lambda \nabla W(\mathbf{r})$, then the linearized equations of motion relative to the surface equilibrium point can be written as

$$\begin{aligned}
\ddot{\xi} &= -2\omega_y \dot{\zeta} + 2\omega_z \dot{\eta} - V_{xx}\xi - V_{xy}\eta - V_{xz}\zeta - \mu N_x \Theta_x \xi - \mu N_x \Theta_y \eta - \mu N_x \Theta_z \zeta \\
\ddot{\eta} &= -2\omega_z \dot{\xi} + 2\omega_x \dot{\zeta} - V_{xy}\xi - V_{yy}\eta - V_{yz}\zeta - \mu N_y \Theta_x \xi - \mu N_y \Theta_y \eta - \mu N_y \Theta_z \zeta \\
\ddot{\zeta} &= -2\omega_x \dot{\eta} + 2\omega_y \dot{\xi} - V_{xz}\xi - V_{yz}\eta - V_{zz}\zeta - \mu N_z \Theta_x \xi - \mu N_z \Theta_y \eta - \mu N_z \Theta_z \zeta
\end{aligned}. \tag{10}$$

The characteristic equation of the surface equilibrium is then changed into

$$\begin{vmatrix} \lambda^2 + V_{xx} + \mu N_x \Theta_x & -2\omega_z \lambda + V_{xy} + \mu N_x \Theta_y & 2\omega_y \lambda + V_{xz} + \mu N_x \Theta_z \\ 2\omega_z \lambda + V_{xy} + \mu N_y \Theta_x & \lambda^2 + V_{yy} + \mu N_y \Theta_y & -2\omega_x \lambda + V_{yz} + \mu N_y \Theta_z \\ -2\omega_y \lambda + V_{xz} + \mu N_z \Theta_x & 2\omega_x \lambda + V_{yz} + \mu N_z \Theta_y & \lambda^2 + V_{zz} + \mu N_z \Theta_z \end{vmatrix} = 0. \tag{11}$$

An equilibrium point is non-degenerate if all the eigenvalues are nonzero.

### 2.3 Number and Stability of Equilibria

Denote $\boldsymbol{\varepsilon} = \begin{bmatrix} \xi & \eta & \zeta \end{bmatrix}^T$, $\mathbf{M} = \mathbf{I}_{3\times 3} = \begin{pmatrix} 1 & 0 & 0 \\ 0 & 1 & 0 \\ 0 & 0 & 1 \end{pmatrix}$, $\mathbf{G} = \begin{pmatrix} 0 & -2\omega & 0 \\ 2\omega & 0 & 0 \\ 0 & 0 & 0 \end{pmatrix}$, and

$\mathbf{K}_\Theta = \mu \begin{pmatrix} N_x \Theta_x & N_x \Theta_y & N_x \Theta_z \\ N_y \Theta_x & N_y \Theta_y & N_y \Theta_z \\ N_z \Theta_x & N_z \Theta_y & N_z \Theta_z \end{pmatrix}$. Then the characteristic equation (11) can be

rewritten as

$$\mathbf{M}\ddot{\boldsymbol{\varepsilon}} + \mathbf{G}\dot{\boldsymbol{\varepsilon}} + \mathbf{K}_V \boldsymbol{\varepsilon} + \mathbf{K}_\Theta \boldsymbol{\varepsilon} = 0. \tag{12}$$

Let $\mathbf{K}_S = \dfrac{\mathbf{K}_\Theta + \mathbf{K}_\Theta^T}{2}$, then we have the following corollary about the linear stability of the surface equilibria.

**Corollary 1.** If the matrix $\mathbf{K}_V$ and $\mathbf{K}_S$ are positive definite, the equilibrium point of the grain on the surface of the irregular body is linearly stable.

**Proof:**

Let $\mathbf{K}_R = \dfrac{\mathbf{K}_\Theta - \mathbf{K}_\Theta^T}{2}$, then $\mathbf{K}_\Theta = \mathbf{K}_S + \mathbf{K}_R$, $\mathbf{K}_S$ is symmetry while $\mathbf{K}_R$ is antisymetry. Let the Lyapunov function be

$$V_{Lyap} = \frac{1}{2}\left(\dot{\boldsymbol{\varepsilon}}^T \mathbf{M} \dot{\boldsymbol{\varepsilon}} + \boldsymbol{\varepsilon}^T \mathbf{K}_V \boldsymbol{\varepsilon} + \boldsymbol{\varepsilon}^T \mathbf{K}_S \boldsymbol{\varepsilon}\right).$$

Then

$$\dot{V}_{Lyap} = \dot{\boldsymbol{\varepsilon}}^T \left(\mathbf{M}\ddot{\boldsymbol{\varepsilon}} + \mathbf{K}_V \boldsymbol{\varepsilon} + \mathbf{K}_S \boldsymbol{\varepsilon}\right) = -\dot{\boldsymbol{\varepsilon}}^T \mathbf{G} \dot{\boldsymbol{\varepsilon}} - \dot{\boldsymbol{\varepsilon}}^T \mathbf{K}_R \dot{\boldsymbol{\varepsilon}}.$$

Considering the matrices $\mathbf{M}$, $\mathbf{K}_V$, and $\mathbf{K}_S$ are positive definite, and $\dot{\boldsymbol{\varepsilon}}^T \mathbf{G} \dot{\boldsymbol{\varepsilon}} = 0, \dot{\boldsymbol{\varepsilon}}^T \mathbf{K}_R \dot{\boldsymbol{\varepsilon}} = 0$, thus $V_{Lyap} = \dfrac{1}{2}\left(\dot{\boldsymbol{\varepsilon}}^T \mathbf{M} \dot{\boldsymbol{\varepsilon}} + \boldsymbol{\varepsilon}^T \mathbf{K}_V \boldsymbol{\varepsilon} + \boldsymbol{\varepsilon}^T \mathbf{K}_S \boldsymbol{\varepsilon}\right) > 0$, and the equilibrium point on the surface is linearly stable. □

The corollary implies that the zero solution of the linearized system is Lyapunov stable, and the surface equilibrium point of system Eq. (12) is linearly stable.

Let $\lambda_k(E_j)$ be the $k$ th eigenvalue of the $j$ th equilibrium. Then we have the following theorem about the number of the equilibria.

**Theorem 1.** If there are totally $N$ surface equilibria, eigenvalues of all the surface equilibria satisfy $\sum_{j=1}^{N}\left[\operatorname{sgn}\prod_{k=1}^{6}\lambda_k(E_j)\right] = \sum_{j=1}^{N}\left[\operatorname{sgn}(\det(\mathbf{K}_V + \mathbf{K}_\Theta))\right] = const$.

**Proof:**

Denote $\dot{\boldsymbol{\varepsilon}} = \boldsymbol{\delta}$, then Eq. (12) can be expressed as

$$\begin{bmatrix} \dot{\boldsymbol{\varepsilon}} \\ \dot{\boldsymbol{\delta}} \end{bmatrix} = \begin{pmatrix} \mathbf{0} & \mathbf{I}_{3\times 3} \\ -\mathbf{M}^{-1}(\mathbf{K}_V + \mathbf{K}_\Theta) & -\mathbf{M}^{-1}\mathbf{G} \end{pmatrix} \begin{bmatrix} \boldsymbol{\varepsilon} \\ \boldsymbol{\delta} \end{bmatrix}. \tag{13}$$

Let $\boldsymbol{\chi} = \begin{bmatrix} \boldsymbol{\varepsilon} \\ \boldsymbol{\delta} \end{bmatrix}$, then Eq. (13) can be simplified into

$$\dot{\boldsymbol{\chi}} = \mathbf{f}(\boldsymbol{\chi}) = \mathbf{A}\boldsymbol{\chi}, \tag{14}$$

where

$$\mathbf{A} = \begin{pmatrix} \mathbf{0} & \mathbf{I}_{3\times 3} \\ -\mathbf{M}^{-1}(\mathbf{K}_V + \mathbf{K}_\Theta) & -\mathbf{M}^{-1}\mathbf{G} \end{pmatrix}. \tag{15}$$

Obviously,

$$\det \mathbf{A} = \det(\mathbf{K}_V + \mathbf{K}_\Theta). \tag{16}$$

Let

$$h(\mathbf{r}) = \left( \frac{\partial V}{\partial x} \quad \frac{\partial V}{\partial y} \quad \frac{\partial V}{\partial z} \right)^T + \left( \mu\Theta N_x \quad \mu\Theta N_y \quad \mu\Theta N_z \right)^T, \tag{17}$$

then $\dfrac{dh}{d\mathbf{r}} = \mathbf{K}_V + \mathbf{K}_\Theta$, define $\deg(f, \Omega, (0, 0, 0))$ as the topological degree of f on $\Omega$ with respect to zero, where $\Omega$ is the definition domain of $f$. Using the topological degree theory (Mawhin and Willem 1989), we have

$$\sum_{j=1}^{N} \left[ \operatorname{sgn} \prod_{k=1}^{6} \lambda_k(E_j) \right] = \deg(f, \Omega, (0, 0, 0)).$$

Considering $\prod_{k=1}^{6} \lambda_k(E_j) = \det \mathbf{A}$, we have

$$\sum_{j=1}^{N} \left[ \operatorname{sgn} \prod_{k=1}^{6} \lambda_k(E_j) \right] = \sum_{j=1}^{N} \left[ \operatorname{sgn}(\det \mathbf{A}) \right] = \sum_{j=1}^{N} \left[ \operatorname{sgn}(\det(\mathbf{K}_V + \mathbf{K}_\Theta)) \right] = const. \tag{18}$$

□

Denote $\operatorname{ind}(E_j) = \operatorname{sgn} \prod_{k=1}^{6} \lambda_k(E_j)$, where $E_j$ is the $j$-th surface equilibrium, $\lambda_k(E_j)$ is the $k$-th eigenvalue of $E_j$. Then, if the surface equilibrium $E_j$ is degenerate, we have $\text{ind}(E_j) = 0$; if the surface equilibrium $E_j$ is non-degenerate, $\text{ind}(E_j) = 1$ or $\text{ind}(E_j) = -1$.

**Corollary 2.** The number of non-degenerate surface equilibria of the grain on the surface of the irregular body can only change in pairs.

**Corollary 3.** A degenerate surface equilibrium point can i) disappear; ii) change to even number of non-degenerate surface equilibria; iii) change arbitrary number of degenerate surface equilibria; iv) change to even number of non-degenerate surface equilibria and arbitrary number of degenerate surface equilibria.

For example, a degenerate surface equilibrium point can change to 2 non-degenerate surface equilibria, or change to 4 non-degenerate surface equilibria and 3 degenerate surface equilibria.

**Corollary 4.** The number of non-degenerate surface equilibria is an even number. It can be 0, 2, 4, 6, 8, …, etc.

**Proof for Corollary 2 and 3:**

The surface equilibria satisfy $\sum_{j=1}^{N}\left[\text{sgn}\prod_{k=1}^{6}\lambda_k(E_j)\right] = const$. If $E_j$ is a degenerate surface equilibria, there are at least one eigenvalue equal zero, so $\prod_{k=1}^{6}\lambda_k(E_j) = 0$. If $E_j$ is a non-degenerate surface equilibria, $\text{sgn}\prod_{k=1}^{6}\lambda_k(E_j) = 1$ or -1. Suppose that there are odd number of non-degenerate surface equilibria, then for each equilibrium, $\text{sgn}\prod_{k=1}^{6}\lambda_k(E_j)$ vary from 1 to -1 or 0, or from -1 to 1 or 0, which make the value of $\sum_{j=1}^{N}\left[\text{sgn}\prod_{k=1}^{6}\lambda_k(E_j)\right]$ non-conservation. So non-degenerate surface equilibria can only change in pairs. One gets Corollary 2.

For a degenerate surface equilibria, we have $\prod_{k=1}^{6} \lambda_k(E_j) = 0$, suppose it change to even (including 0, the number of 2K) number of non-degenerate surface equilibria and arbitrary number (including 0, the number is L) of degenerate surface equilibria, and K of the non-degenerate surface equilibria have index $\mathrm{ind}(E_j) = 1$, the other K of the non-degenerate surface equilibria have index $\mathrm{ind}(E_j) = -1$. Then the total number of index vary from zero to 1×K+(-1)×K+0×L=0, which have no influence to the conservative quantity $\sum_{j=1}^{N}\left[\mathrm{sgn}\prod_{k=1}^{6}\lambda_k(E_j)\right] = const$. One gets Corollary 3.

□

**Proof for Corollary 4:**

Consider the shape of the irregular body; let the body vary to a homogeneous sphere. Then for the rotating homogeneous sphere, it has 2 surface equilibria, the south pole and the north pole. If the rotating speed is zero, it has 0 surface equilibrium. Using this conclusion and Corollary 2, we know that the number of non-degenerate surface equilibria is always an even number.    □

**3. Hopping motion on the Surface**

If the grain hops on the surface of the irregular minor celestial bodies, the motion includes impact motion, orbital motion as well as motion on the surface. In this section, we discuss the motion equations of orbital motion and impact motion.

## 3.1 Orbital Motion

When the grain hops from the surface, the motion equation (Jiang et al. 2014; Jiang and Baoyin 2014) of the grain relative to the irregular body is

$$\ddot{\mathbf{r}} = -2\boldsymbol{\omega} \times \dot{\mathbf{r}} - \frac{\partial V(\mathbf{r})}{\partial \mathbf{r}}. \tag{19}$$

Each time the grain hops, the grain motion is governed by the small body orbital dynamics. The Jacobian integral (Scheeres et al. 1996; Scheeres 2006; Jiang and Baoyin 2014; Chanut et al. 2014) can be expressed as

$$H = \frac{1}{2}\dot{\mathbf{r}}\cdot\dot{\mathbf{r}} - \frac{1}{2}(\boldsymbol{\omega}\times\mathbf{r})\cdot(\boldsymbol{\omega}\times\mathbf{r}) + U(\mathbf{r}). \tag{20}$$

## 3.2 Impact on the Surface

Consider a grain impact on a flat surface (Bellerose and Scheeres 2008), the velocity after impact can be expressed as

$$\begin{cases} v_n = \varepsilon_n v_0 \sin(\alpha_0) \\ v_t = v_0 \cos(\alpha_0) - \mu(1+\varepsilon_n) v_0 \sin(\alpha_0) \end{cases}, \tag{21}$$

where $\varepsilon_n$ and $\mu$ are restitution and friction coefficients, respectively; $v_0$ is the speed before impact, $\alpha_0$ is the angle between the velocity before impact and the flat surface, $v_n$ and $v_t$ represent the normal and tangential speed after impact. After the first impact, the grain will hop and reach the second impact. After several hops, the grain will not hop anymore and remain static on the surface, where the settling time was show to be (Bellerose and Scheeres 2008; Liu et al. 2013) Eq. (22)

$$T = \frac{2v_{0n}}{g_{0n}(1-\varepsilon_n)}, \tag{22}$$

where $v_{0n}$ is the normal speed before the first impact, $g_{0n}$ is the gravitational acceleration at the first impact point.

If the surface is irregular, the motion trajectory and the time for the grain to be static can't be written in an analytical expression; the numerical method is needed to calculate these parameters. Now we present the orbital and surface motion of hopping grains on an irregular asteroid's surface using a combination of an *N*-body gravity algorithm and the soft-sphere discrete element method (Peters and Džiugys 2002; Bierwisch et al. 2009; Schwartz et al. 2012). Given that the impact process (Tardivel and Scheeres 2014) follows soft collisions, both elastic effect and damping mechanism should be included in the analyses of the contact process. To study the coupled orbital and surface dynamics for the hopping grains, a complete simulation must include the mutual interactions between the grains and the asteroid's surface and the gravitational potential of the irregular asteroid. Yu and Baoyin (2015) proposed a numerical method based on the polyhedral method (Werner and Scheeres 1997) to simulate the motion of migrating grains on the surface of asteroids. In this study, a different method based on the *N*-body method is applied, which has its own advantages in the modeling of unsmooth surface. An asteroid can be modeled as a rigid gravitational aggregate of smaller, indestructible spheres (Miyamoto et al. 2007; Richardson 1995). A moderate resolution model of monodisperse particles on a dense hexagonal grid is applied to represent the asteroid and calculate its gravitational potential, as shown in Fig. 2a (hereafter, gravity model). Ting (1993) gave a detailed introduction to this method. In brief, the dynamical behavior of each body can be described by the equations of motion of rigid-bodies. The motion of the mass center of a body obeys Newton's equations, and is integrated using the leapfrog algorithm, while the rotation is derived from the rigid-body Euler equations, and is solved through a high-order time-adaptive Runge-Kutta method. The orientation of the aggregate is described by the four quaternions. From the perspective of asteroid structure, this model can generally reflect the loose and porous characteristics of the majority of asteroids between ~100 m and ~100 km in size (Richardson et al. 2002), which has been successfully used to study the collision process and aggregation dynamics in space (e.g., Matthews and Hyde 2004; Michel and Richardson 2013). The gravitational force of the asteroid and the surface equilibria of a smooth grain on the asteroid are calculated by the polyhedron model. However, the polyhedron models have edges, which make the calculation of the surface motion imprecise. So we use the N-body model to calculate the surface motion. In addition, to resolve the surface motion of grains on an asteroid, a package of fine particles with fixed radii is used to represent the irregular surface of the asteroid, which can accurately depict the characteristics of the asteroid's surface, as shown in Fig. 2b and Fig. 2c (hereafter, surface model). Soft-sphere discrete element method is used to compute contact forces for grains that collide with the surface particles.

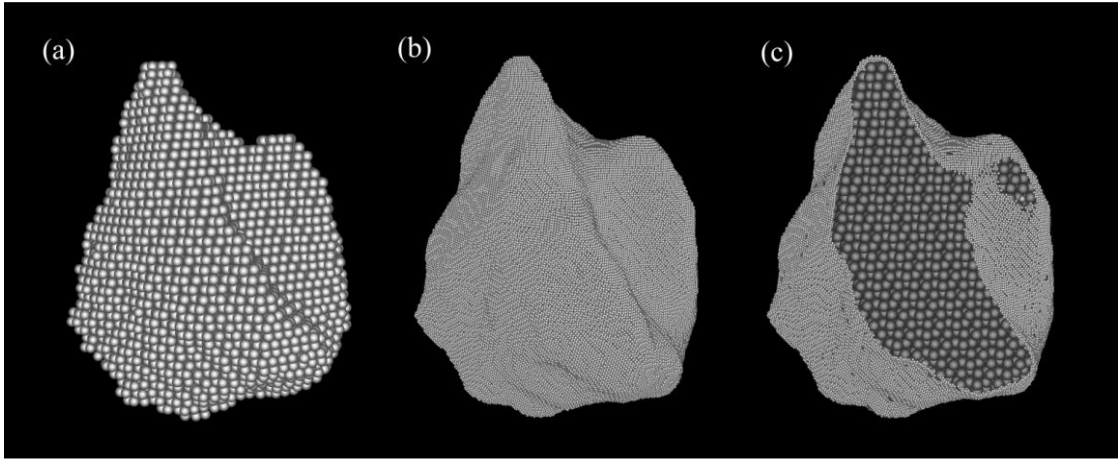

Figure 2. An illustration of the use of a high resolution model comprised of particles to represent an irregular celestial body: (a) interior particle distributions for computing the attraction of asteroid; (b) surface particle distributions for solving the surface migrating motion of grains; (c) cutaway view of the asteroid body. The asteroid 6489 Golevka is employed as an example.

Using this approach, the force acting on each grain can be given by

$$\mathbf{F}_g = \sum_{j=1}^{N} \mathbf{F}_{gj}^{(g)} + \sum_{j=1}^{N_C} \mathbf{F}_{gj}^{(c)}, \tag{23}$$

where $N$ is the total number of particles in the asteroid's gravity model, and $N_c$ is the coordination number of the grain, $\mathbf{F}_{gj}^{(g)}$ and $\mathbf{F}_{gj}^{(c)}$ are the gravitational pull and contact force (if it exists) acting on the grain, respectively.

## 4. Application to the Surface Motion on Asteroid 6489 Golevka

We choose asteroid 6489 Golevka to calculate the surface motion because 6489 Golevka has an irregular shape, and the physiognomy of this asteroid has convex region, concave region and flat surface, which is useful for analysis of the surface equilibria and surface motion behavior affected by the irregular shape. Figure 3 shows the height of surface and gravitation on the surface. The heights of surface mean the distance between the mass center of 6489 Golevka and the surface points. The maximum height is 394.68m and the minimum height is 159.26m. The maximum

gravitation is 2.00416×10⁻⁴N and the minimum gravitation is 1.428542×10⁻⁴N.

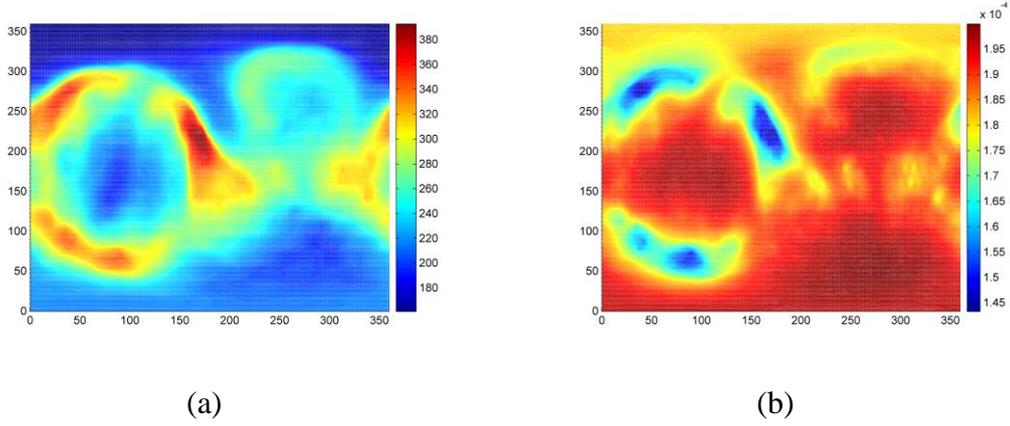

(a) (b)

Figure 3. The height of surface and gravitation on the surface of asteroid 6489 Golevka. (a) the height surface; (b) the gravitation on the surface. The axis units are degree. The colorbar unit in (a) is m while the colorbar unit in (b) is N.

**4.1 Surface Equilibria of a Smooth Grain**

In this section, we compute surface equilibria of a smooth grain. Figure 4 shows the positions of surface equilibria of a smooth grain on the surface of 6489 Golevka, compared with the equilibria in the potential of 6489 Golevka. Table 1 presents the positions and eigenvalues of surface equilibria shown in Figure 4. The eigenvalues are computed by Eq. (5) in Section 2.1. There are 4 equilibria in the potential of 6489 Golevka, all of them are outside the body of 6489 Golevka (Jiang et al. 2014). For a smooth grain, there are 6 surface equilibria on 6489 Golevka, all of them are unstable. From Figure 4, one can see there are 4 surface equilibria, which have the similar azimuth angles and vertical angles with the 4 equilibria in the potential. The total number of surface equilibria of a smooth grain is an even number, which is 6.

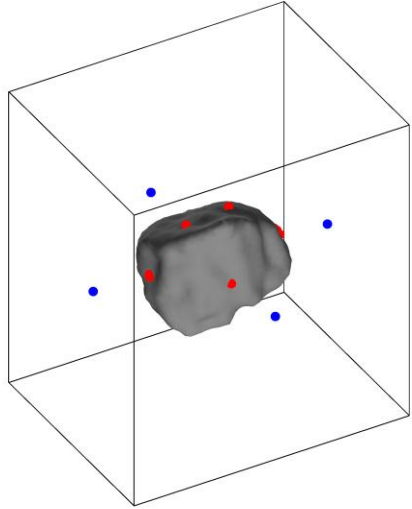

Figure 4. Positions of surface equilibria and equilibria in the potential of 6489 Golevka

Table 1. Positions and eigenvalues of surface equilibria

|   | Positions(m) | Eigenvalues($s^{-1}$) |
| --- | --- | --- |
| 1 | (-95.158, 344.034, 585.848) | $\pm 2.0221 \times 10^{-4}$, $\pm 1.48903i \times 10^{-4}$ |
| 2 | (-7.1560, -322.006, 591.375) | $\pm 1.8983 \times 10^{-4}$, $\pm 2.0323i \times 10^{-4}$ |
| 3 | (-22.227, -835.952, -36.537) | $\pm 1.4014 \times 10^{-4}$, $\pm 1.1750i \times 10^{-4}$ |
| 4 | (-789.896, 39.753, -40.270) | $\pm 2.7273 \times 10^{-4}$, $\pm 2.4725i \times 10^{-4}$ |
| 5 | (-39.384, 1007.679, 12.123) | $\pm 8.6076 \times 10^{-5}$, $\pm 1.0406i \times 10^{-4}$ |
| 6 | (597.525, -17.469, -50.793) | $\pm 3.5071 \times 10^{-4}$, $\pm 3.7148i \times 10^{-4}$ |

## 4.2 Modeling of Hopping Motion on the Surface

### 4.2.1 Numerical Method

In this study, a combination of an N-body gravity algorithm and the soft-sphere discrete element method (SSDEM) is adapted to simulate the hopping motion of a single particle on the surface of an irregular asteroid. Following the contact model introduced in Cundall and Strack (1979), a linear-spring dashpot model is used to describe the normal contact force $\vec{F}_n$ and the tangential sliding resistance $\vec{F}_t$. We also consider the stick-slip effect (Fahnestock and Scheeres, 2009) in this model by setting the coefficient of friction $\mu$ to 1, corresponding to a surface angle of friction ~ 45°. By this mean, there is a limitation for the tangential component of the contact force, and the sliding motion will be activated when the tangential force exceeds the given threshold. In brief, the contact force between two particles is given by

$$F_n = -k_n \xi + C_n u_n,$$
$$F_t = \min(k_t D, \mu_s |F_n|), \qquad (24)$$

which depend on the spring constants, $k_n$ and $k_t$, and the plastic damping parameter, $C_n$ (which is related to the normal coefficient of restitution, $\varepsilon_n$). The variable $\xi$ is the mutual compression of these two particles, and $D$ is the total tangential elongation during this collision. The dashpot force is linearly proportional to the normal relative speed $u_n$. A similar DEM model has been validated through comparison with laboratory experiments (e.g., Schwartz et al., 2014) and has been successfully used to study the dynamics of rubble-pile asteroids (Sánchez and Scheeres, 2012; Zhang et al., 2015). In our study, the values of $k_n$ and integration timestep are set to ensure the maximum of interparticle penetration is always less than 1% of the smallest particle radius in the simulation. The tangential spring constant $k_t \sim (2/7)k_n$. The normal coefficient of restitution is set to 0.5, based on the typical value for rough rocks.

In general, the gravity model with $N \sim 10^4$ is capable of accurately resolving the irregular gravitational field (Richardson et al. 2002). In order to resolve the details of surface motion of grains, the surface model needs a relatively high number of particles (e.g., for a km-size asteroid, its surface model can be built by about $10^5$ m-size particles). However, since the first summation in Eq. (23) only considers the particles in the gravity model and the coordination number is generally less than ten, the proposed method is both accurate and efficient.

**4.2.2 Hopping Motion above Three Special Areas on Asteroid 6489 Golevka**

Take asteroid 6489 Golevka (Stooke 2002; Jiang et al. 2014) as an example. The rotational period of asteroid 6489 Golevka is 6.026 h, its estimated bulk density is $2.7 \text{ g} \cdot \text{cm}^{-3}$, and its three-axes lengths are $0.35 \times 0.25 \times 0.25$ km (Mottola et al. 1997). As shown in Fig. 2, the gravity model includes 9450 mascons on a 10 meter grid, and the surface model includes 263286 particles of 1 m in radius. The grain is modeled as a single sphere with mass $m_g = 2 \times 10^3$ kg and a diameter of 4 m, corresponding to a typical spacecraft. Choose three different impact areas, including the flat surface, concave region, and convex region. The height to release the grain is stochastically chosen in the range of [10, 100] m, and the initial velocity of the grain is zero relative to the body-fixed frame of the irregular body. We use the method introduced above to compute the motion of the grain. Figure 5-7 gives the motion of grains after they are released above these three areas. The motion trajectory of the grains and velocity are expressed in the irregular body's body-fixed frame. Table 2 presents the time parameters for the motion of the grain released above these three areas. Since Equation (22) was derived based on the assumption that the surface is flat and the numerical simulation was conducted using a real irregular asteroid shape, there should be some differences between the results gained from these two methods. For example, the net static times (static time – first impact time) in the simulations are 5626 s, 6256 s, and 16100 s for the case of flat surface, concave region, convex region, respectively.

Taking the value of $g_{0n}$ as $10^{-4}$ m/s$^2$, $\varepsilon_n$ as 0.5, the corresponding surface motion times calculated from Equation (22) are 5300 s, 5490 s and 3130 s for the case of flat surface, concave region, convex region, respectively. For the flat surface case, Equation (22) gives a good estimation of surface motion time. For the other case, due to the complex hopping motion and irregular asteroid shape, the surface motion time cannot be deduced from Equation (22), which is consistent with our derivation.

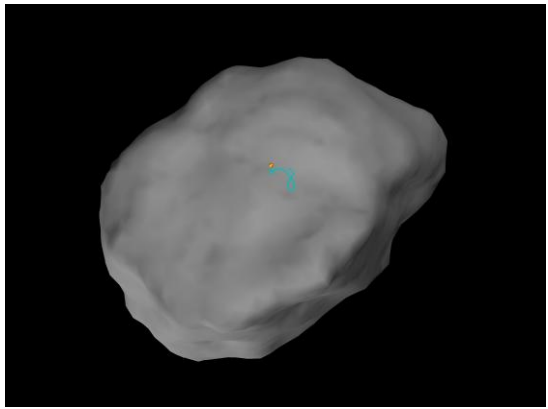

a．Motion trajectory of the grain

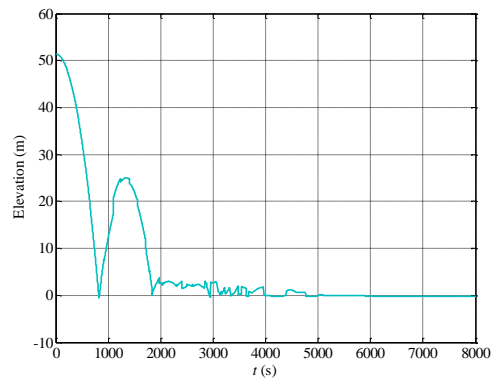

b．Height change of the grain

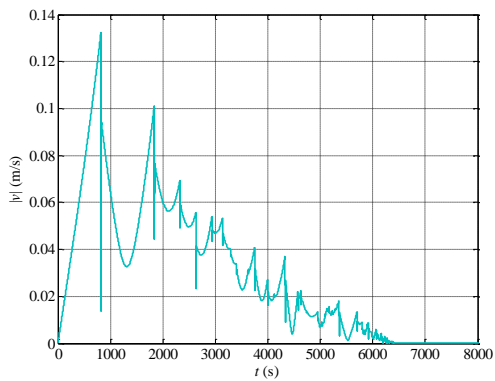

c. The speed of the grain relative to the body

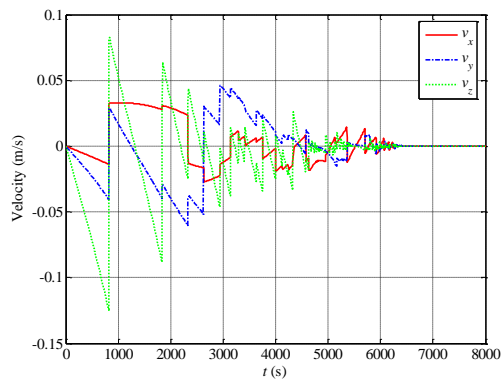

d．The velocity of the grain, red, blue, green represent the velocity component in the x, y, z axis, respectively

Figure 5. Motion of the grain after it is released above the flat surface of the irregular asteroid 6489 Golevka

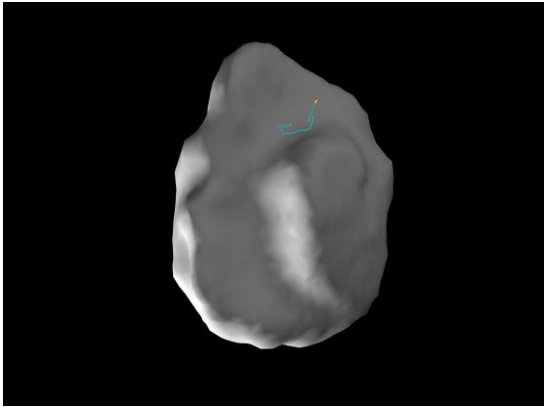

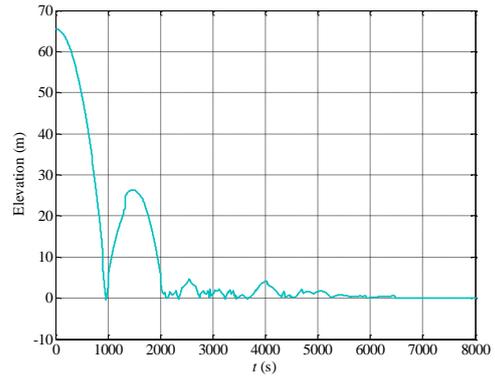

a．Motion trajectory of the grain

b．Height change of the grain

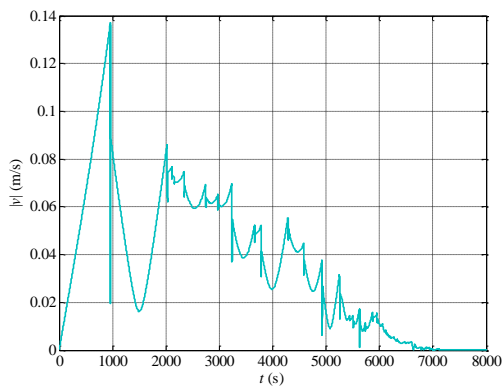

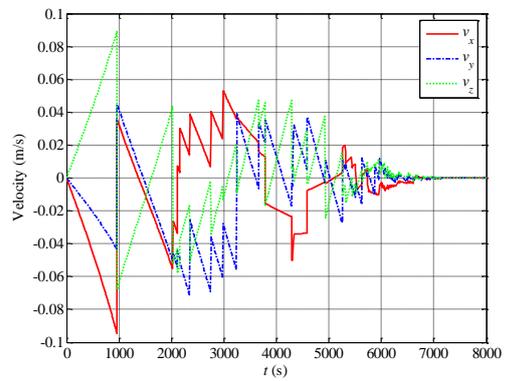

c．The speed of the grain relative to the body

d．The velocity of the grain, red, blue, green represent the velocity component in the x, y, z axis, respectively

Figure 6. Motion of the grain after it is released above the concave region of the irregular asteroid 6489 Golevka

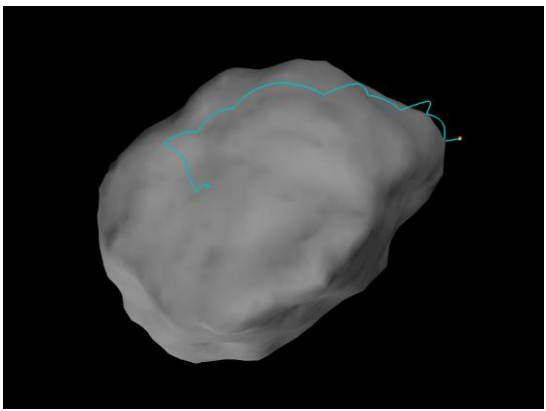

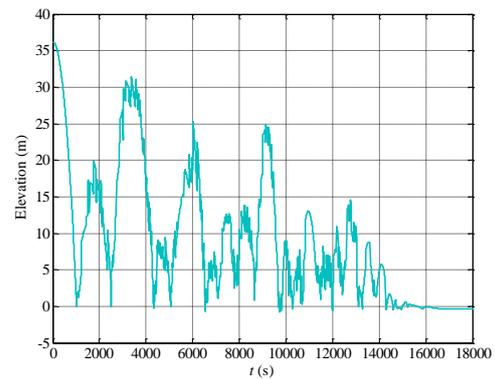

a．Motion trajectory of the grain

b．Height change of the grain

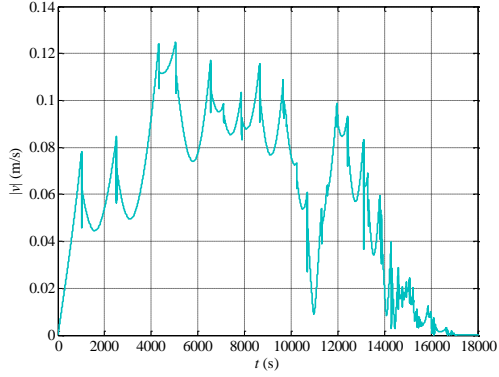
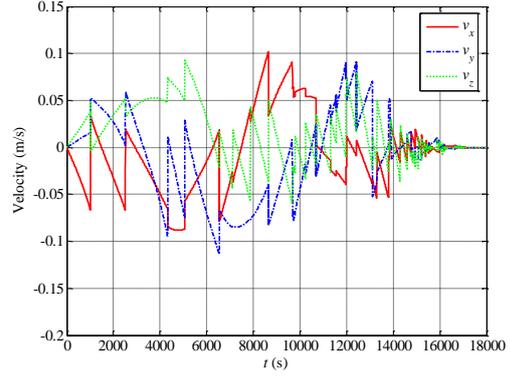

c．The speed of the grain relative to the body

d．The velocity of the grain, red, blue, green represent the velocity component in the x, y, z axis, respectively

Figure 7. Motion of the grain after it is released above the convex region of the irregular asteroid 6489 Golevka

Table 2. Time parameters

| Impact area | First impact time (s) | Relaxed time (s)† | Static time (s)‡ |
| --- | --- | --- | --- |
| Flat surface | 820 | 1820 | 6446 |
| Concave region | 960 | 2013 | 7216 |
| Convex region | 1020 | 14270 | 17120 |

From the motion trajectories of the grains in Fig. 5 to Fig. 7, we can see that after the release, the grain will first be in the orbital motion, then hop on the surface, and then be in the orbital motion again. After several hops, the grain will motion on the surface of the asteroid. From Fig.5-7, one can see that the grain released above the convex region needs significant longer time of motion before being stationary on the surface equilibrium of the irregular asteroid 6489 Golevka. However, the grain released above the flat surface and the concave region have significantly shorter trajectory before being static on the surface equilibrium. Besides, the heights of the grain released above the flat surface and the concave region will decrease rapidly, and the heights of the grain released above the convex region will decrease slowly.

---

† After this moment, the distance between the grain and the surface of the asteroid no longer larger than the sum of the radii of the grain and the surface particle.
‡ After this moment, the velocity of the grain in the body-fixed frame of the asteroid is less than $10^{-4}$ m/s.

From the speed and velocity components of the grain relative to the asteroid, we can conclude that each time the speed varies suddenly, the grain impact on the surface of the irregular body. The impact leads to the sudden change of the size and direction of the grain's velocity. After each hopping, the grain reach the maximum height, then the height will reduce and the grain meet the next impact. If the grain has enough energy to overcome the gravitational potential of the irregular body, the grain will hop again; otherwise, the grain will slip or roll on the surface of the body. When the grain is slipping or rolling, if it goes through a cliff, the grain will hop again. Due to the irregular shape of the asteroid, the grain's maximum heights of each hopping are not always descending. The maximum height of next hopping may be bigger than the maximum height of the current hopping. Because of the friction and non-perfect elastic collision, the mechanical energy of the grain will decrease to zero in the end, and the grain will be static on the surface of the irregular celestial body.

**4.2.3 Monte Carlo Simulations of Hopping Motion**

To analyze the results, we choose 100 different grains with different initial positions and calculate the final positions on the surface. The initial heights are randomly selected in the interval of [0, 550]m, and the azimuth angles and the vertical angles are randomly selected in the interval of [0, 360]deg and [0, 180]deg, respectively. The mean height is 221.294m; the maximum height and the minimum height are 512.1m and 3.0m, respectively. The mean first impact time is 3344.5s; the maximum first impact time and the minimum first impact time are 40600s and 200s, respectively. The mean time for the grains to be static on the surface is 4.6099h; the maximum time is 20.931h while the minimum time is 0.8056h. Figure 8 shows the Monte Carlo simulations of the particles motion above the surface of 6489 Golevka. In Figure 8, the cyan big dots represent the initial positions of the grains while the red small dots represent final positions on the surface of asteroid 6489 Golevka. The results show that if the initial positions of the grains are near the equatorial plane of the asteroid, the motion trajectories and motion time of the grains are significant longer than others. For some grains near the equatorial plane that have a bigger initial height, the grains will escape the asteroid.

Red points on the surface in Figure 8 are stable surface equilibria on the unsmooth surface, while red points on the surface in Figure 4 are surface equilibria on the smooth surface or surface equilibria of smooth grains on the unsmooth surface. Compare with these two figures, one can see that the distribution of surface equilibria of smooth case and unsmooth case are different, because the forces of these two cases are different. In the case of smooth, there is no friction force. From Figure 8, one can see that there are no static grains on the convex region of the surface, which implies that there are no stable surface equilibria on the convex region of the surface. On the concave region and flat surface, there are several static grains, which mean that the stable surface equilibria are on the concave region and flat surface.

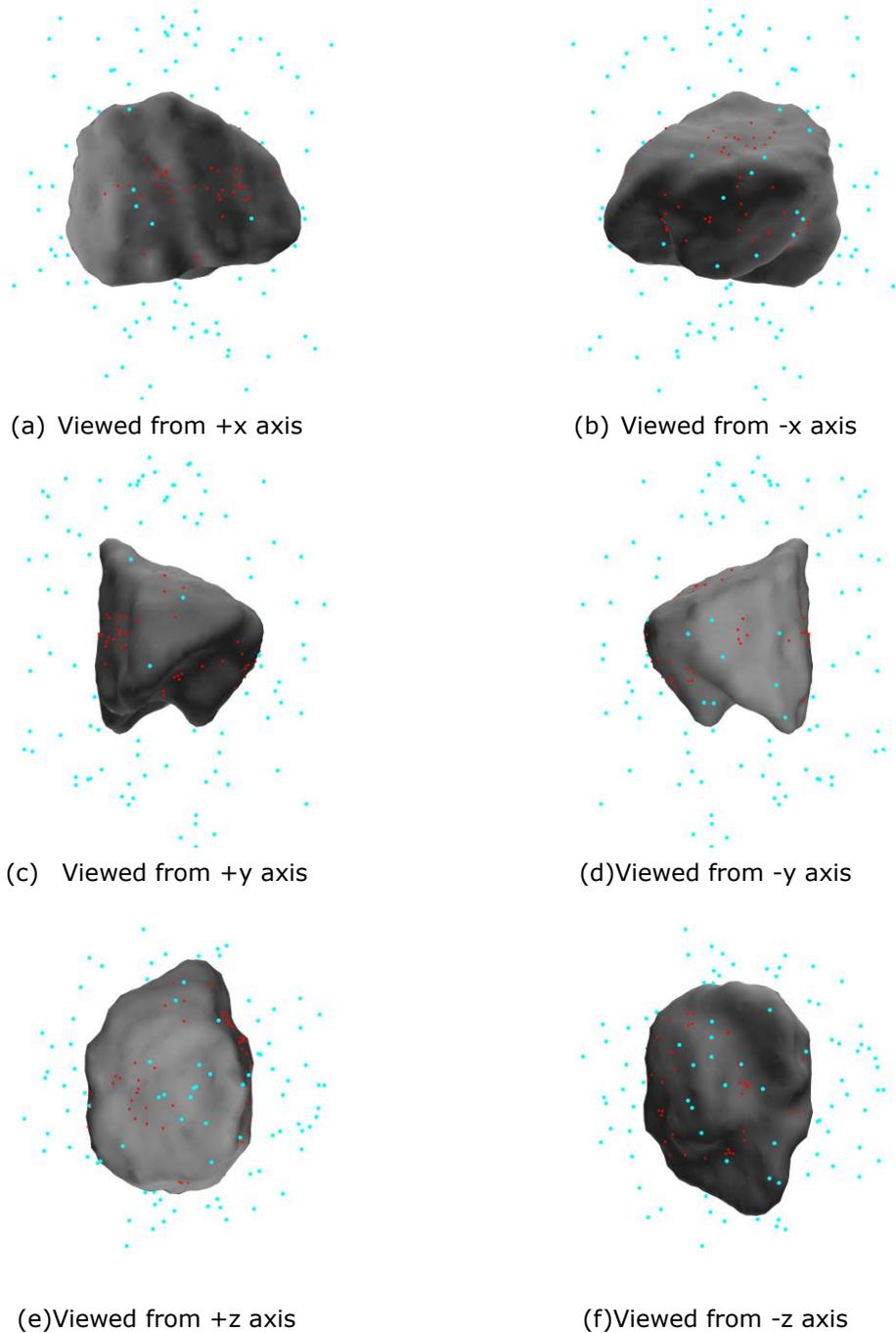

(a) Viewed from +x axis  (b) Viewed from -x axis

(c) Viewed from +y axis  (d) Viewed from -y axis

(e) Viewed from +z axis  (f) Viewed from -z axis

Figure 8. Monte Carlo simulations of the particles motion above the surface of 6489 Golevka, there are 100 grains totally. The cyan big dots represent initial positions while the red small dots represent final positions.

## 5. Conclusions

The surface motions of grains on the irregular celestial bodies are analyzed in this study, which include the following two cases: the grain on the smooth surface and on the unsmooth surfaces. For each case, the linearized motion equations relative to the surface equilibria, the characteristic equations of the surface equilibria, and the sufficient conditions for the linear stability of the surface equilibria are presented. An identity about the eigenvalues of equilibria is presented and proved, the identity implies that the number of non-degenerate surface equilibria is even and the surface equilibria can only change in pairs. The degenerate surface equilibria cannot change to odd number of non-degenerate surface equilibria.

We choose the irregular asteroid 6489 Golevka to calculate the surface equilibria and the motion of a grain released above three different regions: the flat surface, the concave region, and the convex region. There are 6 surface equilibria for a smooth grain on 6489 Golevka. It is found that the grains released above the convex regions on irregular celestial bodies needs significant longer time of motion before being stationary on the surface equilibrium. If the grain is released above the convex region, its height relative to the surface will decrease slowly; if the grain is released above the flat surface or the concave region, its height relative to the surface will decrease rapidly. The grain's speed varies suddenly if and only of the grain impact on the surfaces of the irregular celestial bodies. It is found that the maximum height of next hop may be bigger than the maximum height of the current hop. If the grain does not have enough energy, it will slip or roll on the surface of the body till the grain's mechanical energy decrease to zero and the grain stay static on the surface. Considering the friction force, the Monte Carlo simulations of hopping motions show that there are no stable surface equilibria on the convex region of the surface.

**Acknowledgements**

This research was supported by the National Basic Research Program of China (973 Program, 2012CB720000), the State Key Laboratory Foundation of Astronautic Dynamics (No. 2015ADL02) and the National Natural Science Foundation of China (No. 11372150).

**Appendix A. Supplementary material**

Supplementarydata associated with this article can be found in the online version at http://dx.doi.org/10.1016/j.pss.2016.04.007.

**References**

Asphaug, S. J., Ostro, R. S., Hudson, D. J.: Disruption of kilometer-sized asteroids by energetic collisions. Nature 393, 437-440 (1998)

Bellerose, J., Girard, A., Scheeres, D. J.: Dynamics and control of surface exploration robots on asteroids. In: Hirsch M J, Commander C, Pardalos PM, et al. eds. Optimization and Cooperative Control Strategies. Gainesville: Springer, 2009. 135–150

Bellerose, J., Scheeres, D. J.: Dynamics and control for surface exploration of small bodies. In: Proceedings of AIAA/AAS Astrodynamics Specialist Conference and Exhibit. Honolulu: AIAA, 2008

Bierwisch, C., Kraft, T., Riedel, H., & Moseler, M.: Three-dimensional discrete element models for the granular statics and dynamics of powders in cavity filling. Journal of the Mechanics & Physics of Solids, 57(1), 10–31 (2009)

Chanut, T. G. G., Winter, O. C., Tsuchida, M.: 3D stability orbits close to 433 Eros using an effective polyhedral model method. Mon. Not. R. Astron. Soc. 438(3), 2672-2682 (2014)

Chanut, T. G. G., Winter, O. C., Amarante, A., & Araújo, N. C. S.: 3D plausible orbital stability close to asteroid (216) Kleopatra. Mon. Not. R. Astron. Soc. 452(2), 1316-1327 (2015)

Das, S., Cadirov, N., Chary, S., et al.: Stick–slip friction of gecko-mimetic flaps on smooth and rough surfaces. Journal of The Royal Society Interface, 12(104), 20141346 (2015)

Fahnestock, E. G., Scheeres, D. J.: Binary asteroid orbit expansion due to continued YORP spin-up of the primary and primary surface particle motion. Icarus 201(1), 135-152 (2009)

Guibout, V., Scheeres, D. J.: Stability of surface motion on a rotating ellipsoid. Celest. Mech. Dyn. Astron. 87(3), 263-290 (2003)

Gutiérrez, P. J., Ortiz, J. L., Rodrigo, R.: Effects of irregular shape and topography in thermophysical models of heterogeneous cometary nuclei. Astron. Astrophys. 374(1), 326-336 (2001)

Jewitt, D., Agarwal, J., Li, J., et al.: Disintegrating asteroid P/2013 R3. Astrophys. J. Lett. 784(1), L8 (2014)

Jiang, Y., Baoyin, H.: Orbital mechanics near a rotating asteroid. J. Astrophys. Astron. 35(1), 17-38(2014)

Jiang, Y., Baoyin, H., Li, J., Li, H.: Orbits and manifolds near the equilibrium points around a rotating asteroid. Astrophys. Space Sci. 349(1), 83-106 (2014)

Jiang, Y.: Equilibrium points and periodic orbits in the vicinity of asteroids with an application to 216 Kleopatra. Earth, Moon, and Planets, 115(1-4), 31-44 (2015)

Jiang, Y., Yu, Y., Baoyin, H.: Topological classifications and bifurcations of periodic orbits in the potential field of highly irregular-shaped celestial bodies. Nonlinear Dynam. 81(1-2), 119-140 (2015a)

Jiang, Y., Baoyin, H., Li, H.: Collision and annihilation of relative equilibrium points around asteroids with a changing parameter. Mon. Not. R. Astron. Soc. 452 (4), 3924-3931(2015b)

Jiang, Y., Baoyin, H., Wang, X., et al.: Order and chaos near equilibrium points in the potential of rotating highly irregular-shaped celestial bodies. Nonlinear Dynam. 83(1), 231-252 (2016)

Lagerros, J. S. V.: Thermal physics of asteroids. III. Irregular shapes and albedo variegations. Astron. Astrophys. 325, 1226-1236 (1997)

Liu, X., Baoyin, H., Ma, X.: Dynamics of surface motion on a rotating massive homogeneous body. Sci. China Phys. Mech. 56(4), 818-829 (2013)

Matthews, L. S., Hyde T. W.: Effects of the charge-dipole interaction on the coagulation of fractal aggregates. IEEE Transactions on Plasma Science. 32(2), 586-593 (2004)

Mawhin, J., Willem, M.: Critical point theory and Hamiltonian systems. Springer-Verlag. (1989)

Michel, P., Benz, W., Tanga, P., et al.: Collisions and gravitational reaccumulation: forming asteroids families and satellites. Science 294(5547), 1696-1700 (2001)

Michel P., Richardson D. C.: Collision and gravitational reaccumulation: Possible formation mechanism of the asteroid itokawa. Astron. Astrophys. 554, L1 (2013)

Miyamoto, H., Yano, H., Scheeres, D. J., Abe, S. et al.: Regolith migration and sorting on asteroid itokawa. Science, 316(5827), 1011-1014 (2007)

Mottola, S., Erikson, A., Harris, A. W., et al.: Physical model of near-Earth asteroid 6489 Golevka (1991 JX) from optical and infrared observations. Astron. J. 114(3), 1234-1245 (1997)

Nakamura, T., Noguchi, T., Tanaka, M. et al.: Itokawa Dust Particles: A direct link between S-Type asteroids and ordinary chondrites. Science 333(6046), 1113-1116 (2011)

Oberc, P.: Electrostatic and rotational ejection of dust particles from a disintegrating cometary aggregate. Planet. Space Sci. 45(2), 221-228 (1997)

Peters, B., & Džiugys, A.: Numerical simulation of the motion of granular material using object-oriented techniques. Computer Methods in Applied Mechanics & Engineering, 191(01), 1983–2007 (2002)

Richardson D. C.: A self-consistent numerical treatment of fractal aggregate dynamics. Icarus, 115(2), 320-335 (1995)

Richardson, D. C., Leinhardt, Z. M., Melosh, H. J, et al.: Gravitational aggregates: Evidence and evolution. Asteroids III 1, 501-515 (2002)

Sánchez, D. P., Scheeres, D. J.: DEM simulation of rotation-induced reshaping and disruption of rubble-pile asteroids. Icarus, 218(2), 876-894. (2012)

Stooke, P.: Small body shape models. EAR-A-5-DDR-STOOKE-SHAPE-MODELS-V1.0. NASA Planetary Data System, (2002)

Schwartz, S.R., Richardson, D.C., Michel, P.: An implementation of the soft-sphere dis-crete element method in a high-performance parallel gravity tree-code. Granul. Matter 14(3), 363-380 (2012)

Schwartz, S.R., Michel, P., Richardson, D.C., Yano, H.: Low-speed impact simulations into regolith in support of asteroid sampling mechanism design I: Comparison with 1-g experiments. Planet. Space Sci. 103, 174-183 (2014)

Scheeres, D. J., Ostro, S. J., Hudson, R. S. et al. Orbits close to asteroid 4769 castalia. Icarus 121(1), 67–87 (1996)

Scheeres D. J., Relative equilibria for general gravity fields in the sphere-restricted full 2-body problem. Celest. Mech. Dynam. Astron. 94(3), 317-349(2006)

Tardivel, S., Scheeres, D. J., Michel, P. et al.: Contact Motion on Surface of Asteroid. J. Spacecraft Rockets 51(6), 1857-1871 (2014)

Ting, J. M., Khwaja, M., Meachum, L. R., & Rowell, J. D.: An ellipse-based discrete element model for granular materials. International Journal for Numerical & Analytical Methods in Geomechanics, 17(9), 603–623 (1993)

Werner, R. A.: The gravitational potential of a homogeneous polyhedron or don't cut corners. Celest. Mech. Dyn. Astron. 59(3), 253-278 (1994)

Werner, R. A., Scheeres, D. J.: Exterior gravitation of a polyhedron derived and compared with harmonic and mascon gravitation representations of asteroid 4769 Castalia. Celest. Mech. Dyn. Astron. 65(3), 313-344 (1997)

Yu, Y., Baoyin, H.: Routing the asteroid surface vehicle with detailed mechanics. Acta Mech. Sinica-Prc. 30(3), 301-309 (2014)

Yu, Y., Baoyin, H.: Modeling of migrating grains on asteroid's surface. Astrophys. Space Sci. 355(1), 43-56 (2015)

Zhang, Y., Baoyin, H., Li, J., Richardson, D. C., & Schwartz, S. R.: Effects of orbital ellipticity on collisional disruptions of rubble-pile asteroids. Astrophysics and Space Science, 360(1), 1-16 (2015)